\documentclass[a4paper,fleqn,usenatbib,useAMS]{mnras}

\usepackage{graphicx}	
\usepackage{amsmath}	
\usepackage{amssymb}	
\usepackage{multicol}        
\usepackage{bm}		
\usepackage{pdflscape}	

\usepackage{booktabs} 

\usepackage[normalem]{ulem}

\usepackage{multirow, array} 

\useunder{\uline}{\ul}{}

\usepackage{float} 

\usepackage{hyperref}
\usepackage{tikz}

\definecolor{lime}{HTML}{A6CE39}
\DeclareRobustCommand{\orcidicon}{
	\begin{tikzpicture}
	\draw[lime, fill=lime] (0,0) 
	circle [radius=0.13] 
	node[white] {{\fontfamily{qag}\selectfont \tiny ID}};
	\draw[white, fill=white] (-0.0625,0.095) 
	circle [radius=0.007];
	\end{tikzpicture}
	\hspace{-2mm}
}

\foreach \x in {A, ..., Z}{\expandafter\xdef\csname orcid\x\endcsname{\noexpand\href{https://orcid.org/\csname orcidauthor\x\endcsname}
			{\noexpand\orcidicon}}
}

\usepackage[T1]{fontenc}
\usepackage{ae,aecompl}

\usepackage{newtxtext,newtxmath}

\title[X-rays from CSPN of the Helix Nebula]{Accretion onto WD\,2226$-$210, the central star of the Helix Nebula}

\author[S.~Estrada-Dorado et al.]{S. Estrada-Dorado\thanks{E-mail:\,s.estrada@irya.unam.mx}$^{1}\orcidA$, M.~A.~Guerrero$^{2}\orcidC$, J.~A.~Toal\'{a}\thanks{Visiting astronomer at the IAA-CSIC as part of the Centro de Excelencia
Severo Ochoa Visiting-Incoming programme.}$^{1}\orcidB$, R.~F.~Maldonado$^{1}\orcidE$, V.~Lora$^{3}\orcidF$
\newauthor D.~A.~Vasquez-Torres$^{1}\orcidG$ and Y.-H.~Chu$^{4}\orcidD$\\
$^{1}$Instituto de Radioastronom\'{i}a y Astrof\'{i}sica, Universidad Nacional Aut\'{o}noma de M\'{e}xico, 58090 Morelia, Michoac\'{a}n, Mexico\\ 
$^{2}$Instituto de Astrof\'{i}sica de Andaluc\'{i}a, IAA-CSIC, Glorieta de la Astromom\'{i}a S/N, Granada 18008, Spain\\
$^{3}$Instituto de Ciencias Nucleares, Universidad Nacional Aut\'{o}noma de M\'{e}xico, Apartado postal 73-543, 04510 Ciudad de M\'{e}xico, Mexico\\
$^{4}$Institute of Astronomy and Astrophysics, Academia Sinica, No. 1, Section 4, Roosevelt Road, Taipei 10617, Taiwan
}

\pubyear{2024}

\begin{document}
\label{firstpage}
\pagerange{\pageref{firstpage}--\pageref{lastpage}}
\maketitle

\begin{abstract}
The central star of the Helix Nebula, WD\,2226$-$210 presents enigmatic hard X-ray emission and mid-IR excess. The latter has been attributed to a dusty disk or a cloud-like structure around WD\,2226$-$210 formed from material of Kuiper Belt-like or comet-like objects in highly eccentric orbits. We present here a detailed analysis of multi-epoch {\it Chandra} and {\it XMM-Newton} X-ray observations of WD\,2226$-$210, comparing these to previous \emph{Einstein} and \emph{ROSAT} data. The luminosity of the hard X-ray component of WD\,2226$-$210 has remained basically constant in the decade from 1992 to 2002, with very subtle evidence for variability in timescales of hours. Under the assumption that the X-ray emission from WD\,2226$-$210 is due to accretion of material, an accretion rate of $\dot{M}\approx10^{-10}$ ~M$_\odot$~yr$^{-1}$ is estimated. The origin of the material accreted by WD\,2226$-$210 is uncertain, and can be attributed to the disk-like structure around it or to a sub-stellar donor companion. The accretion rate proposed for the continuous replenishment by bombardment of the mid-IR-emitting structure around WD\,2226$-$210 cannot match that required by the X-ray emission.
\end{abstract}


\section{INTRODUCTION}
\label{sec:intro}

The Helix Nebula is an evolved planetary nebula (PN) located at a relatively short distance \citep[$d \approx$200~pc;][]{Harris2007,BailerJones2021}. Its central star (CS), WD\,2226$-$210 is a hot  H-rich DAO white dwarf (WD) with effective temperature $T_{\rm eff}\sim$110,000~K \citep{Napiwotzki1999,Traulsen2005}. It is expected that its emission should drop steadily at infrared (IR) wavelengths and fade to oblivion at X-ray energies above 0.5~keV. However, the CS of the Helix Nebula has long been known to emit harder X-ray emission even above 1~keV.

The discovery of X-ray emission from WD\,2226$-$210 was done with the {\it Einsten} observatory \citep{Tarafdar1988} and subsequently confirmed by {\it ROSAT} \citep{Leahy1996,Guerrero2000} and \emph{Chandra} \citep[][]{Guerrero2001}. The X-ray spectrum of WD\,2226$-$210 can be described by a combination of a black body component with a temperature of $\approx$130,000~K and a optically thin thermal plasma (see Table~\ref{tab:fluxes}). 
The soft X-ray emission ($E<$0.5~keV) corresponds to the photospheric emission which is better seen in the \textit{ROSAT} data \citep[see fig.~3 of][]{Guerrero2000} and the thermal plasma has a temperature about [7--9]$\times10^{6}$~K that peaks at 0.8~keV \citep[see fig.~2 in][]{Guerrero2001}.

\begin{table*}
\begin{center}
\small
\caption{X-ray Observations of the Helix Nebula and its central star}
\begin{tabular}{lcccccl}
\hline
\multicolumn{1}{l}{X-ray} & 
\multicolumn{1}{c}{Year} & 
\multicolumn{1}{c}{Temperature} & 
\multicolumn{1}{c}{Model} & 
\multicolumn{1}{c}{Obs. Flux} & 
\multicolumn{1}{c}{Acc. Rate} & 
\multicolumn{1}{l}{Reference}\\
\multicolumn{1}{l}{Satellite} & 
\multicolumn{1}{c}{of Obs.} & 
\multicolumn{1}{c}{(10$^{6}$~K)} & 
\multicolumn{1}{c}{} & 
\multicolumn{1}{c}{(10$^{-14}$~erg~cm$^{-2}$~s$^{-1}$)} & 
\multicolumn{1}{c}{(10$^{-10}$~M$_\odot$~yr$^{-1}$)} & 
\multicolumn{1}{l}{}\\
\hline
{\it Einstein} & 1980  & (no spectral fit) & BB & $27\pm 7$ &  & Tarafdar \& Apparao (1988)\\
\hline
{\it ROSAT}    & 1992  & $0.13$ & BB & $19$ &  & Leahy et al.\,(1996) \\
               &       & $4.3$ & plasma & $7.7$ & $2.81$ &                      \\
\hline
{\it Chandra}  & 1999  & (no spectral fit) & BB & (detected )         &  & Guerrero et al.\,(2001)\\
           &       & $$[7--8]$$ & plasma & $5.7$ & $1.72\pm 0.4$ &                     \\
           &   Combination    & $11$ & plasma & $13$ &  $1.10\pm0.2$ &                     \\
\hline
{\it XMM-Newton}& 2002 & (no spectral fit) & BB & (detected) &         & PI: Jansen             \\
               &       &  $7$  & plasma & $8.7$ & $1.72\pm0.1$ &                       \\   
\hline
\label{tab:fluxes}
\end{tabular}
\end{center}
\end{table*}

The origin of the hard X-ray emission from WD\,2226$-$210 was initially attributed to shocked stellar winds filling the Helix Nebula \citep{Leahy1996}, similarly to what is detected in other PNe \citep[see, e.g.,][and references therein]{Chu2001,Ruiz2013,Toala2019}. 
This idea is refuted by the point source nature of the hard X-ray emission at WD\,2226$-$210 determined from \emph{Chandra} observations \citep{Guerrero2001}.  
In addition, \citet{Chu2004a} showed that the {\it FUSE} UV spectrum of this WD does not exhibit any signature of a fast stellar wind. On the other hand, it is known that the hard X-ray emission observed in WDs can be a sign of a stellar companion, with the hard X-rays being was attributed to (i) accretion of material or (ii) coronal X-ray emission from a late-type binary companion \citep{ODwyer2003,Chu2004b,Bilikova2010}.

Temporal variations in the stellar H$\alpha$ profile of WD\,2226$-$210 suggested the presence of an M-type companion \citep{Gruendl2001}, but the near-IR 2MASS photometry ruled it out \citep{ODwyer2003}. Furthermore, {\it TESS} observations revealed photometric variability with a period of $\simeq$2.8~days and small amplitude \citep[$\sim$0.15 per cent;][]{Aller2020,Iskandarli2024} associated with WD\,2226$-$210. 
This variability has been suggested can be attributed to an irradiated Neptune-size (0.021~R$_\odot$ in radius) object in an orbit with inclination $i$=25$^\circ$ \citep{Iskandarli2024}. 
This places the Neptune-size object at an approximated distance of $\sim$6.9~R$_\odot$.

In the mid-IR regime, WD\,2226$-$210 exhibits an excess that can not be explained purely by its stellar emission. This situation is not unique to the CS of the Helix Nebula. IR excess from CS of planetary nebulae suggests the presence of structures such as debris disks \citep[see, e.g.,][and references therein]{Bilikova2012}.  
Similar debris disks are observed around cool chemically-polluted WDs \citep[e.g.,][]{Koester2014,Kilic2012}, usually attributed to the disruption of planetary bodies \citep{Graham1990,Jura2003}, which is the scenario that helps to explain the origin of metal pollution observed in their atmospheres \citep[see][]{Dufour2010}. 
This might be the case of the disk around WD\,2226$-$210, as well as the remains of a circumbinary disk.

\begin{figure*}
\centering
\includegraphics[width=0.5\linewidth]{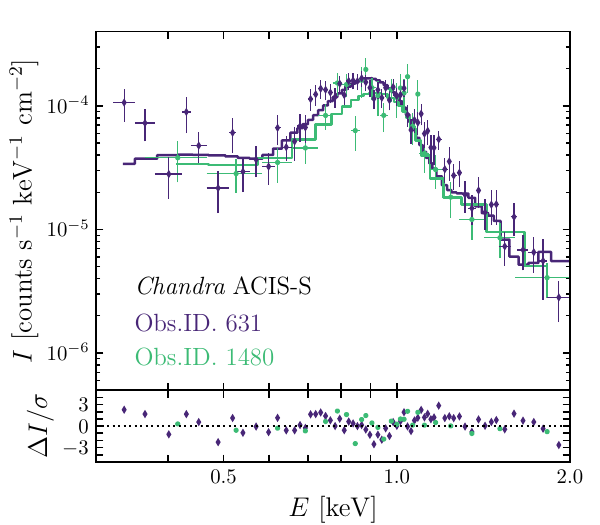}~
\includegraphics[width=0.5\linewidth]{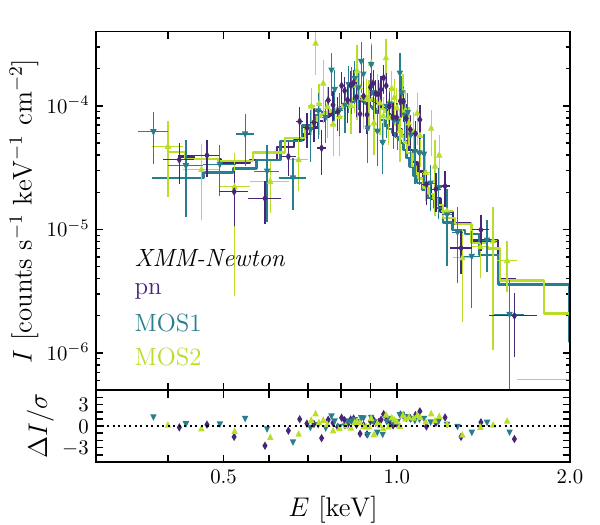}
\caption{Background-subtracted X-ray spectra of WD\,2226$-$210 (the CS of the Helix Nebula).
The left panel shows the {\it Chandra} ACIS-S spectra of the two available observations, while the right panel shows the {\it XMM-Newton} EPIC-pn, MOS1 and MOS2 spectra.
Different colour shades correspond to different observations or instruments as labelled. 
The solid lines show the best-fits obtained with  {\sc xspec} (see Table~\ref{tab:paramters}). The bottom panels represents the deviation of the data from the best-fit models.
}
\label{fig:spec}
\end{figure*}

\citet{Su2007} presented the first model to reproduce the mid-IR excess of WD\,2226$-$210 detected with the {\it Spitzer Space Observatory}. These authors estimated that the structure producing this excess is a dusty disk with  inner and outer radii of $\sim$35 and $\sim$150~AU, respectively. They also suggested that the dust-rich disk might have been formed by material coming from Kuiper Belt–like objects or comets from an Oort-like cloud that survived the post–main sequence evolution.

More recently, \citet{Marshall2023} presented the analysis of WD\,2226$-$210 covering a wide wavelength range (from the mid-IR  to the millimetre) using archival near-IR data and new data acquired from the Stratospheric Observatory for Infrared Astronomy (SOFIA) and the Atacama Large Millimiter/submillimiter Array (ALMA). The non-detection of emission excess at long wavelengths, in conjunction with radiative-transfer models, allowed these authors to reject several hypotheses that may explain the nature and origin of the observed IR excess, including that of a companion to WD\,2226$-$210, a post-AGB circumbinary disk, the presence of an extended planetesimal belt or a disk analogous to the Edgeworth-Kuiper Belt. The favoured scenario to explain the observed dust emission in the circumstellar environment (CE) around the CS of the Helix Nebula is the replenishment of micron-sized dust grains during the pericentre passage of thousands of comets (with masses equivalent to comet Hale-Bopp) having high-eccentricity orbits and whose origin is an Oort cloud-like structure.

With this paper we want to assess the possibility of the accretion of planetary material onto WD\,2226$-$210 as a source powering its hard ($E>$0.5~keV) X-ray emission. For this, we retrieved archival X-ray observations of WD\,2226$-$210 to study possible signatures of variability. This paper is organised as follows. In Section~\ref{sec:obs} we describe the multi-epoch observations. Section~\ref{sec:results} presents our results, which are then discussed in Section~\ref{sec:discussion}. Finally, we summarise our findings in Section~\ref{sec:summary}.

\section{Observations and data preparation}
\label{sec:obs}

\subsection{{\it Chandra} observations}

WD\,2226$-$210 was observed by {\it Chandra} with the Advanced CCD Imaging Spectrometer (ACIS) array on 1999 November 17 and 18 (PI: Y.-H.~Chu) for a total exposure time of 48.7~ks, divided into two exposures of 36.7 and 12.0~ks that correspond to Obs.~ID. 631 and 1480, respectively. 
WD\,2226$-$210 was observed at the aim point of the back-illuminated CCD S7.
These observations were analysed and discussed in \citet{Guerrero2001}.

The {\it Chandra} data were reprocessed here with the Chandra Interactive Analysis of Observations \citep[{\sc ciao}, version~4.14;][]{Fruscione2006} using the \textit{chandra\_repro} task. Time periods of enhanced background emission (i.e., bad periods of time) were detected by inspecting background light curves in the 5.0--10.0~keV. Time periods with count rates above 0.78 ~cnts~s$^{-1}$ were effectively removed, resulting in useful exposure times of 29.98 and 10.04~ks for the Obs.~ID.~631 and 1480, respectively. We extracted ACIS spectra with the \textit{specextract} {\sc ciao} task by defining a circular aperture with a radius of 5~arcsec centred on WD\,2226$-$210. The background spectrum was extracted from an annular region free from background sources with inner and outer radii of 8 and 20~arcsec also centred at the position of WD\,2226$-$210. The individual background-subtracted ACIS-S7 spectra of WD\,2226$-$210 are presented in the left panel of Fig.~\ref{fig:spec}. The background-subtracted ACIS-S count rates of the 631 and 1480 observations resulted in  44.5$\pm$1.2 and 36.2$\pm$1.9~cnts~ks$^{-1}$, respectively.

We also extracted light curves from the two {\it Chandra} observations of WD2226$-$210. This was done by making use of the {\it dmextract} {\sc ciao} task including photons in the 0.5--2.0~keV energy band. 
The resultant light curves are shown in the left panel of Fig.~\ref{fig:ligth-curve}.

\subsection{{\it XMM-Newton} observations}

{\it XMM-Newton} observations of WD\,2226$-$210 were performed as part of the \textit{OM Spectrophotometric Grism Calibration} program (PI: F.\ Jansen). The observations were obtained with the European Photon Imaging Cameras (EPIC) on 2002 November 26 for a total exposure time of 16.8~ks. The total exposure times of the pn, MOS1 and MOS2 cameras are 14.6, 16.3 and 16.3~ks, respectively.

The observations were processed using the Science Analysis Software \citep[{\sc sas}; version 18.0.0;][]{Gabriel2004}. Event files of the pn and MOS cameras were created using the {\it epproc} and {\it emproc} SAS tasks, respectively. After removing time intervals with background count rates in the 10.0--12.0~keV energy range above 0.4~cnts~s$^{-1}$ for EPIC-pn and 0.12~cnts~s$^{-1}$ for the MOS, the net exposure times of the pn, MOS1 and MOS2 cameras resulted in 11.18, 13.93 and 13.93~ks, respectively.

EPIC spectra of WD\,2226$-$210 were extracted with the {\sc sas} tasks {\it evselect} from the three cameras by adopting a circular source aperture of 22~arcsec {\bf in} radius. The background was extracted from an annular region free from background sources with inner and outer radii of 25 and 50~arcsec. The associated calibration matrices were generated with the {\it arfgen} and {\it rmfgen} {\sc sas} tasks. The background-subtracted EPIC spectra of the CS of the Helix Nebula are presented in the right panel of Fig.~\ref{fig:spec}. The background-subtracted count rates for WD~2226$-$210 resulted in 35.0$\pm$1.9 for the EPIC-pn and 13.2$\pm$1.0 for both MOS cameras.

\begin{figure*}
\centering
\includegraphics[height=5.5cm]{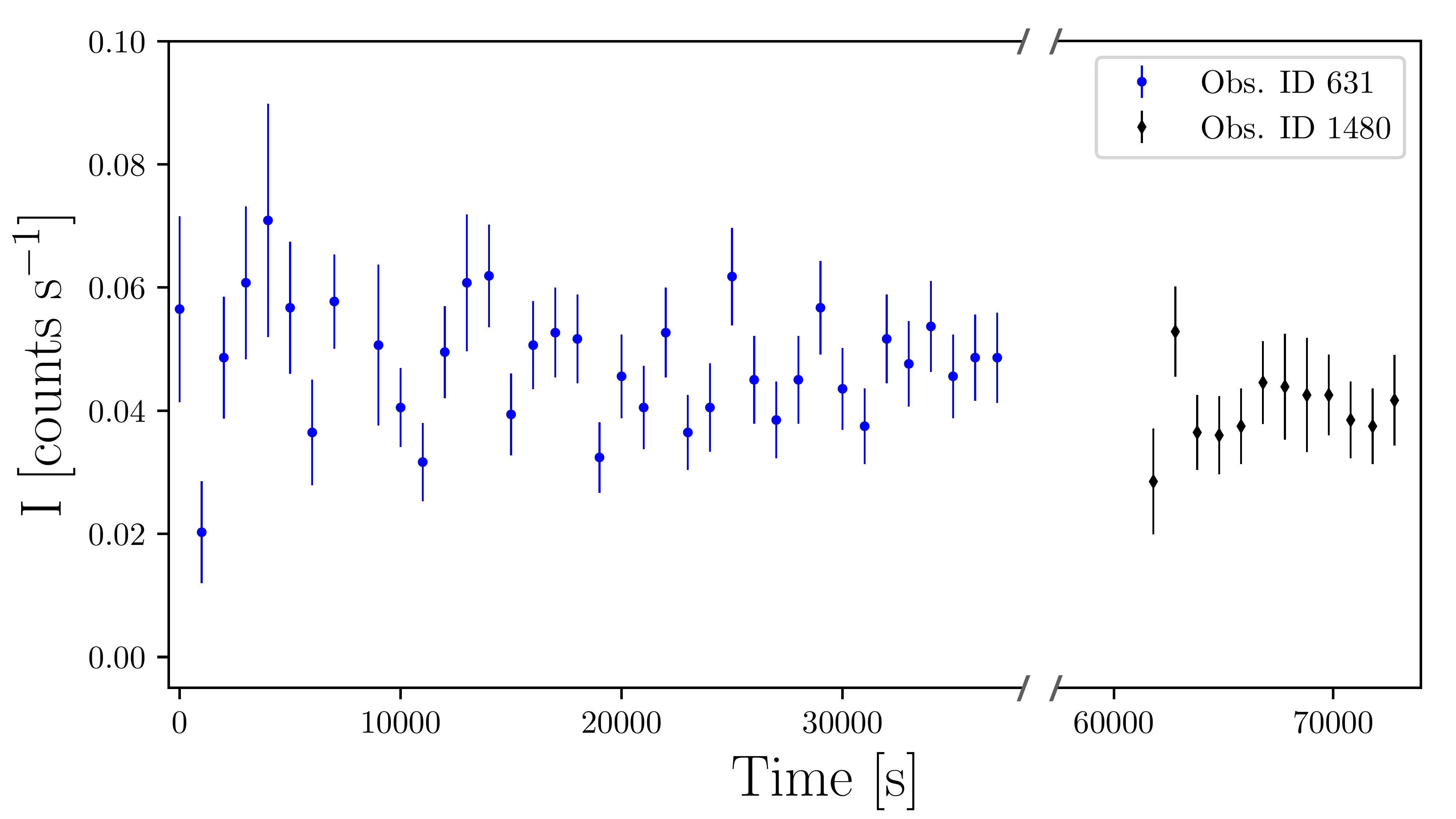}~
\includegraphics[height=5.5cm]{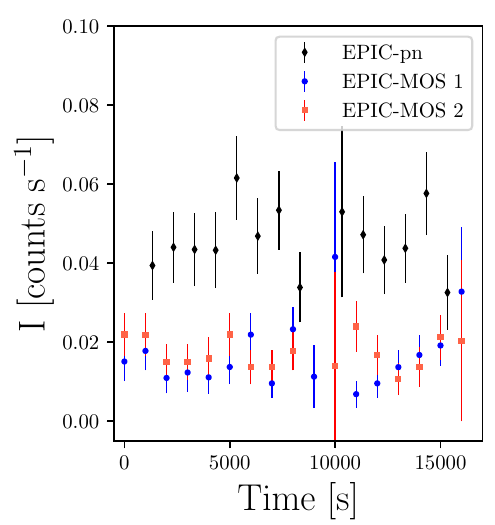}
\caption{
Background-subtracted {\it Chandra} ACIS S7 (top) and {\it XMM-Newton} EPIC (bottom) light curves of WD\,2226$-$210 in the energy range from 0.5 to 2.0 keV. For plotting purposes, the binning in light curves has been set to 1 ks.
}
\label{fig:ligth-curve}
\end{figure*}

Finally, we also extracted light curves from the three EPIC cameras in the 0.5--2.0~keV energy range using the SAS tasks \textit{evselect} and \textit{epiccorr}. These are plotted in the rigth panel of Fig.~\ref{fig:ligth-curve}.

\section{Analysis and results}
\label{sec:results}

\subsection{X-ray spectral Properties}

In order to assess the physical properties of the X-ray emission of the CS of the Helix Nebula, we used the {\sc xspec} package \citep[version 12.10.1;][]{Arnaud1996} to model its \emph{Chandra} and \emph{XMM-Newton} spectra. 
Following the analysis presented by \citet{Guerrero2001}, we adopted an optically thin plasma emission models to fit the spectra. In particular, we adopted the {\it apec} model\footnote{\url{https://heasarc.gsfc.nasa.gov/xanadu/xspec/manual/XSmodelApec.html}} included in {\sc xspec}. We adopted the {\it tbabs} absorption model from \citet{Wilms2000} that takes into account the extinction of X-rays produced by the interstellar medium. Consequently, each fit results in the estimation of the hydrogen column density ($N_\mathrm{H}$) and the plasma temperature of the X-ray-emitting material ($T_\mathrm{X}$).

\begin{table*}
    \begin{center}
    \caption{Best-fit parameters obtained for the models of the X-ray emission detected from the CS of the Helix Nebula, WD\,2226$-$210. The normalisation parameter ($A$) is defined as $A=10^{-14}\int n_\mathrm{H} n_\mathrm{e} dV / 4 \pi d^2$, where $n_\mathrm{H}$ and $n_\mathrm{e}$ are the hydrogen and electron densities, $d$ is the distance and $V$ is the volume of the X-ray-emitting region. Boldface numbers represent fixed values during the spectral fitting procedure.}
    \begin{tabular}{lccccccc} 
    \hline
    \multicolumn{1}{l}{Observation} & \multicolumn{1}{c}{$N_\mathrm{H}$} & \multicolumn{1}{c}{$T_\mathrm{X}$} & \multicolumn{1}{c}{$A$} & \multicolumn{1}{c}{$f_\mathrm{X}$} &\multicolumn{1}{c}{$F_\mathrm{X}$} & \multicolumn{1}{c}{$L_\mathrm{X}$} & \multicolumn{1}{c}{$\chi^2_\mathrm{DoF}$} \\
    \multicolumn{1}{l}{} & \multicolumn{1}{c}{(10$^{20}$~cm$^{-2}$)} & \multicolumn{1}{c}{($kT$)} & \multicolumn{1}{c}{(10$^{-5}$~cm$^{-5}$)} & \multicolumn{1}{c}{(10$^{-13}$~erg~s$^{-1}$~cm$^{-2}$)} & \multicolumn{1}{c}{(10$^{-13}$~erg~s$^{-1}$~cm$^{-2}$)} & \multicolumn{1}{c}{(10$^{29}$~erg~s$^{-1}$)} & \multicolumn{1}{c}{} \\ 
    \hline
    \textit{Chandra} & & & & & & \\ 
    631 & 2.4$\pm$1.3 &0.91$\pm$0.01 & 4.21 & 1.1$\pm$0.1 & 1.2$\pm$0.1 & 5.8$\pm$0.3 & 2.00 \\
    1481 & 1.7$\pm$2.4 & 0.98$\pm$0.03 & 3.74 & 1.0$\pm$0.1 & 1.0$\pm$0.1 & 4.7$\pm$0.5 & 1.52\\
    Combined & 1.9$\pm$1.1 & 0.92$\pm$0.01 & 4.07 & 1.0$\pm$0.1 & 1.2$\pm$0.1 & 5.9$\pm$0.3 & 1.88\\ 
    \hline
    {\it XMM-Newton}\\ 
    pn & 3.7$\pm$1.3 & 0.84$\pm$0.03 & 3.07 & 0.81$\pm$0.4 & 0.9$\pm$0.01 & 4.5$\pm$0.7 & 1.35 \\
    MOS1 & {\bf 3.1} & 0.83$\pm$0.09 & 2.83 & 0.8$\pm$0.1 & 0.9$\pm$0.1 & 4.3$\pm$0.4 &1.03 \\
    MOS2 & {\bf 3.1} & 0.90$\pm$0.08 & 3.05 & 0.8$\pm$0.1 & 0.9$\pm$0.1 & 4.4$\pm$0.6 & 1.22 \\
    EPIC & 1.2$\pm$0.9 & 0.88$\pm$0.02 & 3.72 & 1.0$\pm$0.1 & 1.1$\pm$0.1 & 5.3$\pm$0.7 & 1.49 \\
    \hline 
    \label{tab:paramters}
    \end{tabular}
    \end{center}
\end{table*}

Independent fits were performed on each spectrum extracted from the two different {\it Chandra} observations. The best-fit parameters are listed in Table~\ref{tab:paramters} labelled with their Obs. IDs (631 and 1481). Both models suggest a plasma temperature of $kT\lesssim0.95$ keV with hydrogen column density of $N_\mathrm{H} \approx 2\times10^{20}$ cm$^{-2}$. These models are compared to the observed spectra in the left panel of Fig.~\ref{fig:spec}. In addition, we combined both ACIS-S spectra to produce a single {\it Chandra} ACIS-S spectrum of WD\,2226$-$210 using the {\sc HEASoft} task {\it mathpha}. A fit to this spectrum resulted in similar parameters as those for the individual ACIS-S spectra ($N_\mathrm{H}$=[1.9$\pm$1.1]$\times$10$^{20}$~cm$^{-2}$ and $kT=0.92\pm$0.01 keV). The details are listed in Table~\ref{tab:paramters}, where we also list the normalisation parameter $A$, the observed flux ($f_\mathrm{X}$), intrinsic flux ($F_\mathrm{X}$), and estimated luminosity ($L_\mathrm{X}$) in the 0.3--2.0 keV energy band. The goodness of the fits is evaluated with reduced $\chi^{2}$ statistics ($\chi^{2}_\mathrm{DoF}$), which is also listed in Table~\ref{tab:paramters}. 
The overall fit quality of the simple model adopted here is good, according to the $\chi^{2}_\mathrm{DoF}$ statistics, although its values range between 1.5 and 2.0 and seem to demand more complex emission models, which otherwise can not be well constrained.

Similar analyses were performed on each EPIC spectrum obtained from the {\it XMM-Newton} observations and their results are also listed in Table~\ref{tab:paramters}. We note that the independent fits to each MOS spectrum did not produce reliable $N_\mathrm{H}$ estimations and, thus, we decided to fix this value to 3.1$\times$10$^{20}$~cm$^{-2}$ as reported in the HEASARC $N_\mathrm{H}$ Column Density tool\footnote{\url{https://heasarc.gsfc.nasa.gov/cgi-bin/Tools/w3nh/w3nh.pl}}. The three best-fit models to the EPIC spectra are compared with the observations in the right panel of Fig.~\ref{fig:spec}. For completeness, we also performed a simultaneous fit to the three spectra which resulted in $N_\mathrm{H}$=(1.2$\pm$0.9)$\times$10$^{20}$~cm$^{-2}$ and $kT=$0.88$\pm$0.02 keV, which are very similar to those obtained for the combined {\it Chandra} spectrum (see Table~\ref{tab:paramters}).

\subsection{X-ray time variability}

The spectral analysis presented above indicates that the spectral properties of the X-ray emission detected from WD\,2226$-$210 did not change between 1999 and 2002, the time between the {\it Chandra} and {\it XMM-Newton} observations. It also seems that the fluxes and luminosities of different epochs are consistent, with no significant long-term variation in the X-ray regime, as all flux and luminosity values listed in Table~\ref{tab:paramters} are within error values.
The \emph{Chandra} and \emph{XMM-Newton} X-ray light curves of WD\,2226$-$210 in Fig.~\ref{fig:ligth-curve}, however, suggest a subtle short-term variability.

To assess the possible variable nature of the X-ray emission from WD\,2226$-$210 in the timescales probed by these observations, we used three methods in the analysis of light curves: 
1) the \textit{lcstats} task included in the {\sc HEASoft} tools\footnote{
\url{https://heasarc.gsfc.nasa.gov/ftools}} \citep{Blackburn1995}, which performs a statistical analysis on time series and provides a constant source probability using a Kolmogorov-Smirnov test, 
2) the python routine \textit{pyPeriod} from the PyAstronomy package \citep{Czesla+2019}, which builds the generalised Lomb-Scargle periodogram taking into account the measurement errors and a constant term in the fit of the wave function, 
and 
3) the {\sc pgmuvi} routine \citep{Scicluna2023}, which generates models around the provided data and with Gaussian Process Regression of multi-band time series allows assessing whether there is a variability in the data. 
These three methods were applied to all light curves with time binning values of 0.5, 1.0 and 2.0 ks to search for periods between 2 and 60 ks, which encompass the duration of the longest {\it Chandra} observation.

The \textit{lcstats} task analysis of each observation was used to assess the possible variability of the hard X-ray emission of WD\,2226$-$210. 
Similar analysis was performed on three background point-sources in its vicinity, namely AKARI-FIS-V1\,J2229433$-$204623, [BO98] Helix 341$-$608, and LEDA 835085, to assess possible systematics. 
We found a 50--70 per cent chance of variability of WD\,2226$-$210, depending on the observation, much higher than for the background point-sources.  
In particular, the variability probability of the hard X-ray emission of WD\,2226$-$210 in the {\it XMM-Newton} EPIC-pn light curve is 70 per cent, in contrast with the much lower (less than 10 per cent) chance of variability of point-like sources in its vicinity. 
The much larger likelihood of variability of WD\,2226$-$210 with respect to background point-sources is similar to that found in the analysis of the central star of NGC\,2392, the Eskimo Nebula, and nearby background point-sources, which has been used to support its possible variability \citep{Guerrero2019}.
Based on these results, it is appealing to conclude that WD\,2226$-$210 is X-ray variable.

\begin{figure}
\centering
\includegraphics[width=0.86\linewidth]{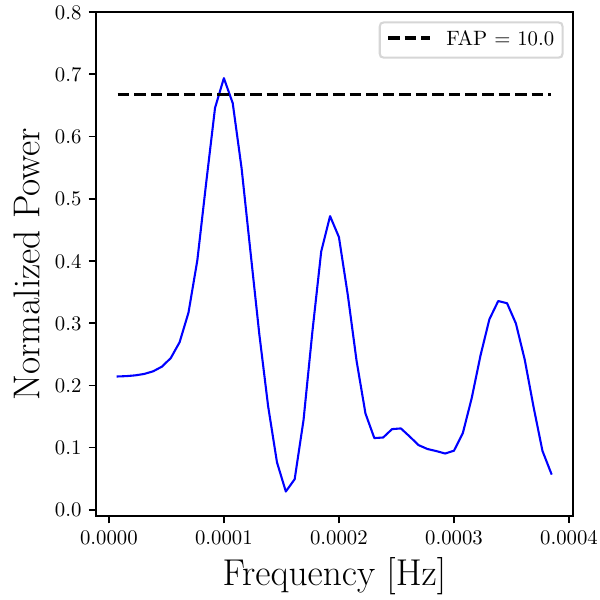}
\caption{
Periodogram of the {\it XMM-Newton} EPIC-pn observations obtained with \textit{pyPeriod}. 
The peaks mark the most feasible periods in frequency units and the dashed line shows the 10\% false alarm probability. See text for details.
}
\label{fig:periodogram}
\end{figure}

The {\it pyPeriod} task and {\sc pgmuvi} routine were then applied to the individual light curves extracted from the {\it XMM-Newton} and {\it Chandra} observations to search for putative variability periods of the hard X-ray emission of WD\,2226$-$210. 
The analysis of all light curves with {\it pyPeriod} results in periods from 2.3 to 10.5~ks. 
This is illustrated by the periodogram of the {\it XMM-Newton} EPIC-pn data shown in Fig.~\ref{fig:periodogram}, where the most probable frequencies result in $\omega_1$=(9.5$\pm$2.3)$\times10^{-5}$, $\omega_2$=(1.9$\pm$0.2)$\times10^{-4}$ and $\omega_3$=(3.5$\pm$0.4)$\times10^{-4}$, that correspond to periods of 10.5$\pm$3.3~ks, 5.2$\pm$0.6 ks, and 2.8$\pm$0.4 ks, respectively\footnote{It seems likely that $\omega_2$ and $\omega_3$ are harmonic frequencies of $\omega_1$, such that $\omega_2 = 2\omega_1$ and $\omega_3 \approx 3\omega_1$.}. 
On the other hand, the {\sc pgmuvi} routine, applying several models looking for the best fit of the data and errors, resulted in periods of 2.4 ks for EPIC-MOS and 10.5 ks for {\it Chandra} and {\it XMM-Newton} EPIC-pn.  
Among all these periods, the \textit{lcstats} task particularly favours the longest ones, with a 70 per cent chance of variability.

\section{Discussion}   
\label{sec:discussion}

The analysis of the X-ray data of the CS of the Helix Nebula presented here shows that there has been no noticeable changes in the flux in time-scales of years. The {\it Chandra} and {\it XMM-Newton} observations cover a time-scale of 3 years and their analyses suggest an averaged X-ray flux in the 0.3--10.0 keV of $f_\mathrm{X}\approx9\times10^{-14}$~erg~cm$^{-2}$~s$^{-1}$, which translates into an intrinsic luminosity of $L_\mathrm{X}\approx5.0\times10^{29}$~erg~s$^{-1}$. Furthermore, flux estimations obtained from previous 1980 {\it Einstein} and 1992 {\it ROSAT} observations suggest similar observed fluxes of (2.7$\pm$0.7)$\times10^{-13}$~erg~cm$^{-2}$~s$^{-1}$ for the photospheric soft and plasma hard components \citep[][]{Tarafdar1988} and 8$\times10^{-14}$~erg~cm$^{-2}$~s$^{-1}$ or the optically thin plasma component \citep[][]{Leahy1996}, respectively. We can safely conclude that no dramatic flux changes have been recorded from WD\,2226$-$210 in the X-ray regime for about 22 years.

Assuming that the X-ray emission we detect from WD\,2226$-$210 is due to accretion, we can use the analytical prescription presented by \citet{Patterson1985} to estimate the mass accretion rate ($\dot{M}_\mathrm{acc}$) from the plasma temperature as:
\begin{equation}
    kT \approx 1.3 \left(\frac{M_\mathrm{WD}}{0.7~\mathrm{M}_\odot}\right)^{3.6} \left( \frac{10^{16}~\mathrm{g}~\mathrm{s}^{-1}}{\dot{M}_\mathrm{acc}} \right),
    \label{eq:acc1}
\end{equation}
\noindent with $M_\mathrm{WD}$ as the mass of the WD. 
The accretion rate derived from a plasma temperature of 0.9 keV, which is consistent to all \emph{Chandra} and \emph{XMM-Newton} spectral fits in Table~\ref{tab:paramters}, is $\dot{M}_\mathrm{acc} \approx 1.2 \times 10^{-10}$~M$_\odot$~yr$^{-1}$.  
This value agrees with the accretion rate derived from the plasma temperature reported for \emph{ROSAT} data in Table~\ref{tab:fluxes} given its large uncertainty, implying a constant accretion rate onto WD\,2226$-$210 in the decade from 1992 to 2002.

The accreted mass onto WD\,2226$-$210 must definitely come from material in its vicinity. To interpret the mid-IR excess of WD\,2226$-$210, \citet{Su2007} proposed the presence of a 35--150 AU in size debris disk around it. The material of this disk would arise from the disruption of Kuiper Belt-like objects or the breakup of comets from an Oort-like cloud. More recently, \citet{Marshall2023} rather proposed that the dust is distributed on a cloud-like structure of material from bodies in highly eccentric orbits released at their periastron passage. They noted that the measured dust mass and lifetime of the dust grains required the disruption of several thousand Hale–Bopp equivalent comets per year to maintain the observed mid-IR excess around the CS of the Helix Nebula. Using the values reported by \citet{Marshall2023}, the replenishment rate of the cloud-like structure is $\approx4\times10^{-12}$~M$_\odot$~yr$^{-1}$. 
This value is about 30 times lower than the accretion rate obtained from the X-ray emission (see above).  
The accretion rate derived from the X-ray emission is found here to be steady, at least in the timescales probed by the available X-ray observations. This implies that there is no direct correlation between the X-ray emission over the years and the cloud-like structure proposed by \citet{Marshall2023}. Thus, other alternatives are worth discussing.

\subsection{A variable source}

Although there is no conclusive evidence for variability of the hard X-ray emission of WD\,2226$-$210 in shorter time scales, our time analysis reveals a 70 per cent chance of variability associated with a period of 10.5 ks.  
This is notably larger than the less than 10 per cent variability chance for sources in its vicinity, which hints at the hard X-ray variability of WD\,2226$-$210.

Some implications on the origin of the hard X-ray emission can still be derived assuming that this 10.5 ks period was real. 
Following the methodology presented in \citet{Chu2021}, we will assume that the variability is due to accretion from a hidden companion on an orbit with a period equal to the X-ray variability period and propose some representative candidates:
{\it i}) a late M6~V companion\footnote{\citet{Gruendl2001} rule out the presence of a companion as late as a M5\,V star.}, {\it ii}) a brown dwarf, and {\it iii}) a Jupiter-like planet. 
For the M6\,V star we adopt a radius of $R$=0.15~R$_\odot$ and mass of $M$=0.1~M$_\odot$, 
for the T-type brown dwarf we adopt $R$=0.1~R$_\odot$ and $M$=0.035~M$_\odot$, and 
for the Jupiter-like planet $R$=0.1~R$_\odot$ and $M$=0.001~M$_\odot$. 
The first step is thus estimating the semi-major orbital separation ($a$) of the components on each system for an orbital period of 10.5~ks.

\begin{table}
\begin{center}
\small
\caption{Separation of the WD and companion ($a_\mathrm{}$), effective Roche Lobe \citep[$r_\mathrm{RL}$;][]{Eggleton1983}, and separation at which the companion is disrupted \citep{Murray1999}.}
\begin{tabular}{lccccc}
\hline
\multicolumn{1}{l}{Companion} & 
\multicolumn{1}{l}{Mass} & 
\multicolumn{1}{l}{Radius} & 
\multicolumn{2}{c}{Period 10.8~ks} & 
\multicolumn{1}{c}{Tidal radius} \\
   & $M$ & $R$  & \multicolumn{1}{c}{$a_\mathrm{}$} & 
\multicolumn{1}{c}{$r_\mathrm{RL}$} & 
\multicolumn{1}{c}{$r_\mathrm{tidal}$} \\
   & (M$_\odot$) & (R$_\odot$) & (R$_\odot$) & (R$_\odot$) &  (R$_\odot$) \\
\hline
M6\,V       & 0.1   & 0.15 R$_\odot$ & 0.94 & 0.22 & 0.38  \\ 
Brown Dwarf & 0.035 & 0.1 R$_\odot$ & 0.92 & 0.16 & 0.36   \\ 
Jupiter     & 0.001 & 0.1 R$_\odot$ & 0.87 & 0.05 & 1.21 \\ 
\hline
\label{tab:radius}
\end{tabular}
\end{center}
\end{table}

The derived semi-major orbital separations\footnote{It is clear that the proposed Jupiter-like planet has a compact orbit around WD\,2226$-$210, a characteristic that has been found to be common from sub-stellar companions orbiting other WDs \citep[][and references therein]{Veras2021}. In such cases, it is accepted that the planet started its evolution at larger initial $a$ values and migrated into close-in orbits due to star-planet interactions \citep[e.g., tidal forces;][]{Villaver2009} and/or by instabilities caused by the presence of other planets \citep[see][and references therein]{Maldonado2021,Maldonado2022}.} are listed in column 4 of Table~\ref{tab:radius}.
For each orbital separation, we estimate the effective Roche Lobe radius ($r_\mathrm{RL}$) using the approximations presented in \citet{Eggleton1983}:
\begin{equation}
    r_\mathrm{RL}/a = \frac{0.49\,q^{2/3}}{0.6\,q^{2/3} + \mathrm{ln}(1 + q^{1/3})},
\end{equation}
\noindent where $q$ is the mass ratio between the companion and the WD. The results are displayed in column 5 of  Table~\ref{tab:radius}, where it is to be noted that only the Jupiter-like planet is capable of filling this effective Roche Lobe and transfer material to the CS of the Helix Nebula.  
Certainly the irradiation of the hot WD\,2226$-$210 can inflate the atmosphere of a companion star, but the predicted radius increase of brown dwarfs \citep{Sainsbury2021} and main sequence stars \citep{Barman2004} in this case is not enough to fill their effective Roche Lobes.  

However, to assess the survival of the companion, we computed the tidal radius $r_\mathrm{tidal}$ for each system, which is the limiting radius at which the orbiting body is disrupted by the tidal forces, defined as \citep[see][]{Murray1999}:
\begin{equation}
r_{\mathrm{tidal}}=\left (\frac{3 \rho_{\mathrm{WD}}}{\rho_{\mathrm{orb}}}\right )^{1/3} R_{\mathrm{WD}},   
\label{eq:radius}
\end{equation}
\noindent where $\rho_{\mathrm{WD}}$ and $\rho_{\mathrm{orb}}$ are the densities of the WD and the orbiting body, respectively, and $R_{\mathrm{WD}}$ is the WD's radius. 
The values of $r_\mathrm{tidal}$ are listed for all systems in column 6 of Table~\ref{tab:radius}.
We found that the Jupiter-like planet is the only disrupted body with the estimated separations.
However, the source of X-ray emission can not be a M6V or a brown dwarf companion because, although they survive the tidal forces, they do not fill their Roche lobes.

Nevertheless, if the density of the planet were not uniform and its core dense enough, it could survive the effect of tidal forces without being completely disrupted.
For example, a planetesimal sized object has been found orbiting the WD SDSS~J1228+1040 at a distance of 0.73 R$_\odot$ from its host WD star. 
This object has been hypothesized to be the iron-rich core of a larger body whose outer layers have been eroded by the tidal forces of the WD \citep{Manser2019}. 
Assuming this same scenario and adopting their upper limit of 7$\times$10$^{24}$~g for the mass of the orbiting iron core, we can estimate that a progenitor planet of this core should have lost its mass at a rate of $5\times10^{-8}$~M$_\odot$~yr$^{-1}$ for the $\approx$20000~yr age of the Helix Nebula to reassemble a planet with the mass of Jupiter. 
For an accretion rate of $10^{-10}$~M$_\odot$~yr$^{-1}$, it implies that the accretion onto WD\,2226$-$210 has an efficiency $\simeq$0.002.

On the other hand, if we calculate the mass that a progenitor orbiting body should have in order to lose mass at the same rate as the accretion rate calculated with the X-ray emission in order to end in as a planetesimal sized body, the initial mass would be $2.4\times10^{27}$~g, which is about half the Earth mass.
In fact, using Eq.~(\ref{eq:radius}) we estimate that any rocky planet similar to the inner planets in the Solar system will survive the tidal disruption if located at $>$0.88 R$_\odot$ given that their averaged density is 3.72~g~cm$^{-3}$. For comparison we note that Venus, Earth and Mars have averaged densities\footnote{\url{https://ssd.jpl.nasa.gov/horizons}} of 5.20, 5.51 and 3.93~g~cm$^{-3}$, respectively.

This scenario is supported by the possibility to have minor planets with mass of $\sim$10$^{27}$ g reaching close distances to the WD if, during the transformation of the main sequence star into the current PN, the planetary system was destroyed. Indeed, numerical simulations have shown that minor planets are prone to acquire high-eccentricity orbits due to planet-planet scattering events and their pericenter may reach star-grazing orbits or even cross Roche-limit distances during the dynamical evolution of multiple-planet systems \citep{Maldonado2021}. Nevertheless, we note that the existence of such a  theorised orbiting body must be accompanied by a debris disk of eroded material \citep[e.g.,][]{Manser2019}, but its existence is ruled out by the present IR observations of WD2226$-$210.

Alternatively the X-ray emission variability period of 10.5 ks can be equalled to the rotation period of the CS, assuming the variability is attributed to a hot spot on its surface. 
The mean rotation periods of isolated WDs, however, seem to be much longer \citep[$\sim$ 35 hrs,][]{Hermes2017}, but it is marginally consistent with those of fast rotation massive or magnetic WDs \citep[e.g.][]{Ferrario2020}.  
These would result from the CS evolution in a binary system, whose effects  are beyond the analysis presented here.

\subsection{No variability}

Leaving aside the probable variability of the hard X-ray emission, during the evolution of the progenitor of WD\,2226$-$210 and the destruction of its associated planetary system, planet-scale collisions generated smaller bodies that could be now accreted by the CS. 
Accretion of this planetary material would be able to power the X-ray emission, which has been indeed
suggested to cause the metal contamination in the atmosphere of cool DZ WDs \citep[e.g.][]{Melis2017,Harrison2018,Hollands2018}.  
If this were to be the case of the CS of the Helix Nebula, it would be expected to find traces of Ca or other metals in its atmosphere, as it is observed in cool metal polluted WDs \citep{Zuckerman2003}. 
However, the high temperature of WD\,2226$-$210 makes it difficult to detect metallic absorption lines in its spectrum due to the fact that the atmospheres of hot WDs are more opaque with respect their cooler counterparts \citep{Badenas2024}. 
Furthermore, if some metals were to be detected, it would be difficult to discern their origin due to the radiative levitation observed in hot stars.  
The metallicity would rather reflect the chemical enrichment caused by the previous stellar evolution, mostly during the AGB phase, and not from accreting material from the circumstellar environment \citep{Vennes1988,Chayer1995}.

Another potential source of accreted material could be the Neptune-sized planet orbiting WD2226$-$210 with a 2.8-day period \citep{Iskandarli2024}, located $\approx6.9$ R$_\odot$ from the WD. Although this planet cannot fill its Roche lobe, photoevaporation might play a significant role in mass loss. For instance, \citet{Gansicke2019} observed that the cooler WD J0914+1914 ($T\mathrm{eff} \approx 27,400$ K) is photoevaporating an ice giant planet located at 15~R$_\odot$.

Although not definite, our findings are in line with recent interpretation of the X-ray emission from other hot WDs. Recently, \citet{Chu2021} suggested that the clearly variable hard X-ray emission from the hot WD KPD\,0005+5106 could be explained by accretion of material from an unseen, orbiting Jupiter-like planet. On the other hand, the X-ray emission from G\,29-38 seems to be dominated by remnant planetary material \citep{Cunningham2022,Estrada-Dorado2023}. In addition, our numerical simulations of a WD accreting material from a mass-losing planet predict significant X-ray emission that is consistent with observations of KPD\,0005+5106 and G\,29-38 \citep{Estrada-Dorado2024}.

\section{Summary}
\label{sec:summary}

We presented the analysis of multi-epoch {\it Chandra} and {\it XMM-Newton} observations of the central star of the Helix Nebula, WD\,2226$-$210. This is a hot WD from which X-rays with energy above 0.5~keV have been reported in the past. 

We found no evidence of variable emission in the timescale of three years between these observations.  
The observed (intrinsic) X-ray flux has remained constant with an averaged value of $f_\mathrm{X}\approx9\times10^{-14}$~erg~cm$^{-2}$~s$^{-1}$ ($F_\mathrm{X} \approx10^{-13}$ erg~cm$^{-2}$~s$^{-1}$) that implies an averaged luminosity of 
$L_\mathrm{X}\approx5.0\times10^{29}$~erg~s$^{-1}$. 
These values seem also consistent with a previous 1992 {\it ROSAT} observation of the hard X-ray emission from WD\,2226$-$210, which suggest a longer steady behaviour in the X-ray regime at least for a decade.

The temporal analysis on short timescales of the {\it Chandra} and {\it XMM-Newton} observations are inconclusive. However, we found a 70 per cent chance of variability from WD\,2226$-$210 with a period of 10.5~ks, against a 10 per cent chance from sources in the vinicity of this WD.
If this variability is to be believed, it is quite different from the 2.77~days ($\sim 240$~ks) optical period.

Assuming that the observed X-ray emission is produced by accretion, we estimate an accretion rate of $\dot{M}_\mathrm{acc}=10^{-10}$ M$_\odot$~yr$^{-1}$. 
We suggest that the most feasible explanation is X-ray emission powered by the accretion of material resulting from a close planet disrupted by the effects of tidal forces rather than from material in a disk or a cloud-like structure around WD\,2226-210 \citep{Su2007,Marshall2023}.

Along with G\,29-38 and KPD\,0005+516, WD\,2226-210 is pushing the limits of our understanding on the production of X-ray emission via accretion of planetary material onto WDs. It is thus imperative to make systematic searches to reveal similar systems. These are the systems that are showing us the remaining signatures of the survival of planetary systems through the evolution of Solar-like stars. 
Future {\it Xrism} observations would allow assessing the abundances of the X-ray-emitting material to inquire into its possible planetary origin.

\section*{Acknowledgements}
The authors thank comments and suggestions from an anonymous referee that helped improving the presentation and discussion of our results.

S.E.-D. thanks support to CONACyT-Mexico for a student grant. J.A.T. acknowledges support from UNAM PAPIIT project IN102324. M.A.G. acknowledges financial support from grants CEX2021-001131-S funded by MCIN/AEI/10.13039/501100011033 and PID2022-142925NB-I00 from the Spanish Ministerio de Ciencia, Innovaci\'on y Universidades (MCIU) cofunded with FEDER funds. R.F.M. acknowledges support from the Programa de Becas posdoctorales of the Direcci\'{o}n General de Asuntos del Personal Acad\'{e}mico of the Universidad Nacional Aut\'{o}noma de M\'{e}xico (DGAPA, UNAM, Mexico). Y.-H.C. acknowledges the support of the grant NSTC 112-2112-M-001-065 from the National Science and Technology Council of Taiwan. This work is based on observations obtained with {\it XMM-Newton}, an European Science Agency (ESA) science mission with instruments and contributions directly funded by ESA Member States and NASA. This research has made use of data obtained from the Chandra Data Archive and software provided by the Chandra X-ray Center (CXC) in the application packages {\sc ciao}. This work has made extensive use of NASA's Astrophysics Data System.

\section*{DATA AVAILABILITY}

The processed data were obtained from the public archives of {\it XMM-Newton} and {\it Chandra}. They will be shared on reasonable request to the corresponding author. The original data sets can be downloaded from public archives.


\end{document}